\documentclass[twocolumn,aps,prb,10pt,amssymb,floatfix]{revtex4-1}
\usepackage{amsmath,amssymb}
\usepackage{hyperref}
\usepackage{braket}
\usepackage{graphicx}
\usepackage{hyperref}
\usepackage{afterpage}

\hypersetup{colorlinks=true, linkcolor=blue, citecolor=blue, urlcolor=blue}

\DeclareMathOperator{\tr}{tr}

 \DeclareMathOperator{\hc}{h.c.}

\renewcommand{\Re}{\operatorname{Re}} 
\renewcommand{\Im}{\operatorname{Im}}

\newcommand{\ci}{\text{i}}

\begin{document}

\date{\today}

\author{Patrick Haughian}
\author{Massimiliano Esposito}
\author{Thomas L. Schmidt}
\email{thomas.schmidt@uni.lu}
\affiliation{Physics and Materials Science Research Unit, University of Luxembourg, 1511 Luxembourg, Luxembourg}

\title{Quantum thermodynamics of the resonant-level model\\ with driven system-bath coupling}

\begin{abstract}
We study nonequilibrium thermodynamics in a fermionic resonant level model with arbitrary coupling strength to a fermionic bath, taking the wide-band limit. In contrast to previous theories, we consider a system where both the level energy and the coupling strength depend explicitly on time. We find that, even in this generalized model, consistent thermodynamic laws can be obtained, up to the second order in the drive speed, by splitting the coupling energy symmetrically between system and bath. We define observables for the system energy, work, heat, and entropy, and calculate them using nonequilibrium Green's functions. We find that the observables fulfill the laws of thermodynamics, and connect smoothly to the known equilibrium results.
\end{abstract}

\maketitle

\section{Introduction}
Thermodynamics has long been central to the understanding and optimization of the performance of work cycles and machines. As a result of recent advances in fabrication technology and measurement techniques, the range of realizable machines now extends down to the size of a single molecule.\cite{vdz12,pek15,ros16,arg17} At this length scale, many of the assumptions underlying the edifice of thermodynamics are no longer valid. Specifically, the paradigm of a nanoscale system coupled to a bath features a variety of subtleties that are not present in macroscopic setups. First, the extent of the contact area between system and bath may be similar to that of the system itself, meaning the details of the coupling become relevant and can no longer be treated in the same general fashion as in the macroscopic case. Second, if the coupling is of sufficient strength, even the distinction between system and bath may itself become blurred.\cite{hae08} Finally, the task of taking into account non-equilibrium effects is much more intricate at the nanoscale and has received a great deal of recent attention. These difficulties are compounded by quantum effects that play no role macroscopically but rise to prominence in small systems. However, even the problem of formulating a microscopic theory of non-equilibrium thermodynamics for a classical system is daunting in itself.\cite{sei16,str17,jar17}

The key to establishing a framework of nonequilibrium quantum thermodynamics is to define quantities that transfer the concepts of system energy, entropy, heat, and work to the nanoscale regime in the context of a given system, while preserving as much generality as possible. To this end, a wide variety of approaches has been pursued, studying setups which roughly fall into the two categories of weak\cite{spo78,breuer_petruccione_book,esp09,kos13,gas15} and general couplings\cite{jar97,jar04,tal07,esp09,cam09,esp10,cam11,kat16,str16,xuu16,kat16_2} between system and bath, respectively. For the case of weak coupling, consistent thermodynamics has been established,\cite{breuer_petruccione_book,esp09,kos13,cue15} but beyond weak coupling, the situation is much less clear: There, the meaning of work and work fluctuations has been understood, but the quest for definitions of system energy and heat remains open.\cite{jar97,tal07,cam11}

Recently, the formalism of nonequilibrium Green's functions has been applied to the question of statistical physics and thermodynamics in paradigmatic quantum systems.\cite{kit10} The advantage of this approach lies in its inherent ability to treat both nonequilibrium and strong-coupling situations, meaning that Green's functions can readily provide a wide range of candidates for thermodynamic definitions. The subjects of these studies are variants of the resonant level model, consisting of an electronic level coupled to metallic leads, under the influence of a drive protocol. This constitutes a minimal description of a quantum dot coupled to source and drain electrodes and driven by means of ac gate voltages. Electronic transport in the time-dependent resonant level model and its extensions has been studied for several years,\cite{keeling08,schmidt08,riwar09,haughian16,haughian17} but more recently its thermodynamic properties have come into the spotlight. Several sets of thermodynamic definitions have been proposed in this way, with varying ranges of validity.\cite{bru16,esp15,och16,gal15,lud13} In particular, it has proved challenging to find appropriate generalizations of corresponding equilibrium quantities,\cite{esp15} and to incorporate drive protocols and coupling structures of general form.\cite{och16,bru16}

Our work considers a resonant level model in the spirit of Ref.~[\onlinecite{bru16}], where the electron level is subjected to a time-dependent drive and coupled to a single lead, which we consider in the wide-band limit. To arrive at a more realistic model for experiments,\cite{dutta17} we extend the existing models by in addition allowing for a time-dependent coupling between system and bath, and show that it admits an analytical solution in terms of Green's functions. These solutions give rise to nonequilibrium thermodynamic quantities, which connect smoothly to their equilibrium counterparts, and obey the laws of thermodynamics in the quasi-adiabatic limit. In doing so, we give a definition of the heat current which differs from those considered in Ref.~[\onlinecite{gal15}] and thus resolve the apparent inconsistency caused by time-dependent coupling.

The paper is structured as follows: In Sec.~\ref{sec_model}, we introduce the resonant level model and its solution in the presence of time-dependent parameters. Next, we use this solution to define thermodynamic quantities in Sec.~\ref{sec_firstlaw}, and demonstrate the first law of thermodynamics in our model. We proceed in Sec.~\ref{sec_equilibrium} by confirming that the adiabatic limit of our definitions matches established equilibrium results. In Sec.~\ref{sec_secondlaw} we perform an expansion in derivatives of the drive protocol, from which we conclude that our definitions are compatible with the second law of thermodynamics up to second order in drive velocities. We compare this expansion with exact numerical results in Sec.~\ref{sec_numerics}. Finally, we summarize our findings and compare to related results in the literature, in Sec.~\ref{sec_discussion}.

\section{Resonant level model}
\label{sec_model}
\begin{figure}[!t]
     \includegraphics[width=0.99\columnwidth]{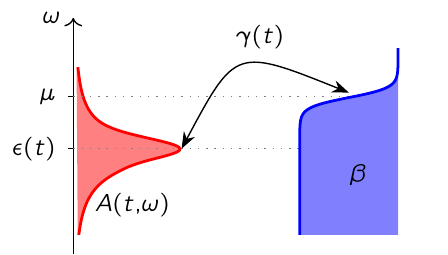}
 \caption{Resonant level model of a driven single-electron quantum dot at energy \(\epsilon(t)\), with time-dependent tunnel coupling \(\gamma(t)\) to a single metallic lead at inverse temperature \(\beta\), with chemical potential \(\mu\). The coupling results in broadening of the dot electron level with profile \(A(t,\omega)\), see Eq.~\eqref{number}.}
\label{fig_model}
\end{figure}

We study a model Hamiltonian for a single electronic level coupled to a fermionic lead,
\begin{align}
\label{hamiltonian}
    H(t) &= H_{\text{D}}(t) + H_{\text{B}} + H_{\text{T}}(t), \\
    H_\text{D}(t) &= \epsilon(t) d^\dag d, \notag \\
    H_\text{B} &= \sum_k \epsilon_k c_k^\dag c_k, \notag \\
    H_\text{T}(t) &= \sum_k \gamma(t)d^\dag c_k + \hc, \notag
\end{align}
where $H_\text{D}(t)$, $H_\text{B}$, and $H_\text{T}(t)$ are the dot, lead, and tunneling Hamiltonians, respectively. Here, \(d^\dag\) and \(d\) denote the creation and annihilation operators for the dot electron and fulfill the fermionic commutation relation \(\{d,d^\dag\}=1\). Analogously, the operators \(c^\dag_k\) and \(c_k\) are associated with the lead electrons, with the index \(k\) enumerating the lead modes. In the absence of the tunneling term, we assume thermal equilibrium in the lead, thus imposing \(\langle c^\dag_k c_q\rangle_0=\delta_{kq}f(\epsilon_k)\), where \(f(\epsilon_k)=[1+\text{e}^{\beta(\epsilon_k-\mu)}]^{-1}\) denotes the Fermi-Dirac distribution at inverse temperature \(\beta\) with chemical potential \(\mu\), and the subscript 0 denotes expectation values taken with respect to the quadratic Hamiltonian \(H_\text{D}(t)+H_\text{B}\). Both the dot energy \(\epsilon(t)\) and the dot-lead coupling strength \(\gamma(t)\) are subject to time-dependent drive, and no assumption is made regarding the magnitude of \(\gamma(t)\). In this way, the Hamiltonian in Eq.~\eqref{hamiltonian} combines non-equilibrium physics and potentially strong coupling between system and bath, and hence features several of the challenges inherent in the attempt to formulate quantum thermodynamics. The schematics of the model are visualized in Fig.~\ref{fig_model}.

\begin{figure}[t!]
     \includegraphics[width=0.99\columnwidth]{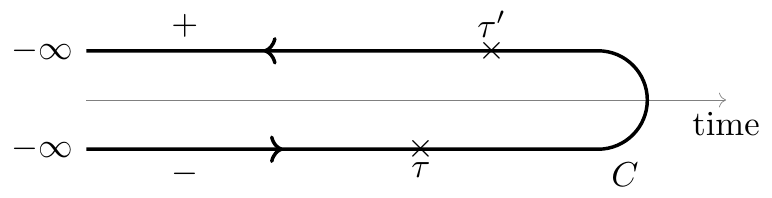}
 \caption{Keldysh integration contour \(C\) with times \(\tau\) and \(\tau'\), running from \(-\infty\)  to \(+\infty\) in the lower half plane, before returning to \(-\infty\) in the upper half plane.}
\label{fig_keldysh}
\end{figure}

In the following, we describe the dynamics of the resonant level model of Eq.~\eqref{hamiltonian} in terms of nonequilibrium Green's functions,\cite{hau94} which can be calculated analytically. Specifically, we consider the tunneling Hamiltonian as an interaction term and use a perturbation series to capture the renormalization of dot properties as a consequence of this interaction. The starting point for this procedure is given by the bare dot Green's function
\begin{align}
D_0(\tau,\tau')\equiv-\ci \left\langle \mathcal{T}_C d(\tau)d^\dag(\tau')\right\rangle_0.
\end{align}
Owing to the non-equilibrium nature of the problem, the times \(\tau\) and \(\tau'\) are defined on the Keldysh contour \(C\) as seen in Fig.~\ref{fig_keldysh}, with \(\mathcal{T}_C\) as the corresponding path ordering symbol.

In terms of real-valued times \(t\) and \(t'\), the Green's function takes on the matrix structure
\begin{align}
\check{D}_0(t,t')=\begin{pmatrix}D_0^{\text{R}}(t,t')&D_0^{\text{K}}(t,t')\\0&D_0^\text{A}(t,t')\end{pmatrix},
\end{align}
with \(D_0^{\text{R}}\), \(D_0^{\text{A}}\), and \(D_0^{\text{K}}\) denoting the retarded, advanced and kinetic Green's functions, respectively. We also introduce the lesser Green's function \({D_0^{-+}(t,t')=\ci \left\langle d^\dag(t') d(t)\right\rangle_0}\). In the absence of coupling, the Heisenberg equation of motion leads to the following time-evolution of the dot operators,
\begin{align}
d(t)=d(0)\exp\left[-\ci \int_0^{t}\text{d}s\epsilon(s)\right],
\end{align}
which allows one to determine the bare dot Green's functions.

\begin{figure}[t!]
     \includegraphics[width=0.99\columnwidth]{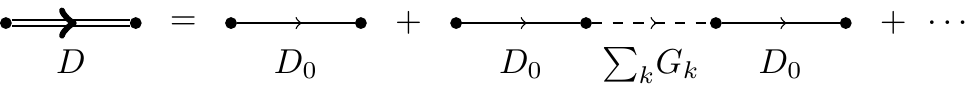}
 \caption{Feynman diagrams contributing to the exact dot Green's function from Eq.~\eqref{greens_expansion}.}
\label{fig_diagrams}
\end{figure}

The exact Green's function in the interaction picture is defined by
\begin{align}
\label{greens_expansion}
D(\tau,\tau')=-\ci \left\langle\mathcal{T}_C d(\tau)d^\dag(\tau')\text{e}^{-\ci \int_C\text{d}\sigma H_\text{T}(\sigma)}\right\rangle_0,
\end{align}
where the time evolution of all operators is governed by the unperturbed Hamiltonian $H_\text{D} + H_\text{B}$. Since we do not assume the coupling to be small, all orders of the resulting series must be taken into account. However, the diagrams arising in this way are all of the linear structure depicted in Fig.~\ref{fig_diagrams}, so the series can be resummed and yields the Dyson equation
\begin{align}
\label{dyson}
D(\tau,\tau')&=D_0(\tau,\tau')\notag\\
&+\int_C\text{d}\sigma\text{d}\sigma' D_0(\tau,\sigma) \Sigma(\sigma,\sigma') D(\sigma',\tau').
\end{align}
Thus, the consequences of coupling to the leads are fully quantified by the self-energy
\begin{align}
\label{sigma}
\Sigma(\sigma,\sigma')=\gamma(\sigma)\gamma^*(\sigma')\sum_{k}G_{0,k}(\sigma,\sigma'),
\end{align}
where \(G_{0,k}(\sigma,\sigma')=-\ci \langle\mathcal{T}_C c^{\phantom\dag}_k(\sigma)c^{\dag}_k(\sigma')\rangle_0\) denotes the bare lead Green's function. To calculate this function analytically, we assume the wide-band limit, i.e., a linear spectrum $\epsilon_k = v_F k$ with infinite bandwidth. The wide-band limit constitutes an excellent approximation at temperatures less that the Fermi energy of the bath, and leads to a constant density of states $\rho_0 = L/(2 \pi v_F)$, where \(L\) denotes the spatial extent of the lead. The retarded self energy is then given by
\begin{align}
\label{sigmaret}
\Sigma^\text{R}(s,s')&=
-\ci \Gamma(s)\delta(s-s'),
\end{align}
where we introduced the tunneling linewidth \({\Gamma(s)=\pi\rho_0|\gamma(s)|^2}\), which we assume to be strictly positive. Importantly, the wide-band limit produces a delta-shaped \(\Sigma^\text{R}(s,s')\). Similarly, we evaluate the lesser component,
\begin{align}
\label{sigmalesser}
\Sigma^{-+}(s,s')&=
2\pi\ci \rho_0\gamma(s)\gamma^*(s')\int\frac{\text{d}\omega}{2\pi}\text{e}^{-\ci \omega(s-s')}f(\omega),
\end{align}
which depends on the lead distribution \(f(\omega)\).

Eq.~\eqref{dyson} admits an analytical solution: First, we note that we can solve it for the retarded and advanced Green's functions by taking the retarded component of each factor. Then, we proceed by iteratively replacing instances of the exact Green's function \(D^\text{R}\) in Eq.~\eqref{dyson} by the entire right hand side of Eq.~\eqref{dyson}. This leads to an expansion in powers of the self-energy, which sums to
\begin{align}
\label{dret}
D^\text{R}(t,t')=-\ci  \theta(t-t') \text{e}^{-\ci  \int_{t'}^t\text{d}s\epsilon(s)} \text{e}^{-\int_{t'}^t \text{d}s \Gamma(s)}.
\end{align}
The advanced component is then given by \(D^\text{A}(t,t')=D^\text{R}(t',t)^*\). Together with the lesser self-energy, they give rise to the $D^{-+}$ via the Langreth rule,\cite{langreth91,hau94}
\begin{align}
\label{dlesser}
D^{-+}(t,t')=\int_{-\infty}^\infty \text{d}s\ \text{d}s'D^\text{R}(t,s)\Sigma^{-+}(s,s')D^\text{A}(s',t'),
\end{align}
which we simplified by noting that an additional term\cite{hau94}, which is proportional to the dot occupation at the initial time \(t_0\),  drops out since \(t_0\to-\infty\). Using Eqs.~\eqref{sigmalesser} and \eqref{dret}, we thus evaluate the lesser dot Green's function as
\begin{align}
D^{-+}(t,t')=2\ci \int\frac{\text{d}\omega}{2\pi} f(\omega) \text{e}^{-\ci \omega(t-t')} V(t,\omega) V^*(t',\omega),
\end{align}
where we defined the function
\begin{align}
\label{vdef}
V(t,\omega)=\int_{-\infty}^t \text{d}s \sqrt{\Gamma(s)} \exp\left\{\int_s^t \text{d}y \left[\ci \omega-\ci \epsilon(y)-\Gamma(y)\right]\right\}
\end{align}
which encodes the history of the driving protocol \([\epsilon(t),\Gamma(t)]\). By choosing \(t=t'\), this Green's function provides us with the expectation value of the dot particle number, \(N(t)=-\ci D^{-+}(t,t)\),
\begin{align}
\label{number}
N(t)=\int\frac{\text{d}\omega}{2\pi}f(\omega)A(t,\omega),
\end{align}
where we wrote
\begin{align}
\label{adef}
A(t,\omega)=2|V(t,\omega)|^2.
\end{align}
Note that Eq.~\eqref{number} appears as a straightforward generalization of the dot particle number in a stationary system, where \(A(t,\omega)\) would be replaced by the Lorentzian spectral function
\begin{align}
\label{A0def}
A_0(\omega)=\frac{2 \Gamma}{(\omega-\epsilon)^2+\Gamma^2}.
\end{align}
Since it can be shown that in the stationary case \(A\) and \(A_0\) coincide (see App.~\ref{app_limit}), the expression \(A(t,\omega)\) can be viewed as a drive-induced modification of the dot spectral function. However, it bears pointing out that in general \(A\) does not match the definition of the non-stationary spectral function: \({A(t,\omega)\neq-2\Im D^{R}(t,\omega)}\), where \(D^{R}(t,\omega)\) stands for the Wigner transform of the retarded dot Green's function.

We have thus arrived at a fully analytical solution of the resonant level model in the presence of two drives, after taking the wide-band limit. In a way similar to the particle number, expectation values of any other operator on the dot and lead Hilbert spaces can be calculated from the Green's function matrix \(\check{D}(t,t')\).

\section{Thermodynamic definitions and first law}
\label{sec_firstlaw}
In the following, we define thermodynamic quantities for our system in terms of quantum mechanical expectation values and use the Green's functions obtained in the previous section to calculate them. Herein, we require these definitions to be compatible with the laws of thermodynamics, in the presence of dot and coupling drives as well as arbitrary dot-lead coupling strength.

It has been shown previously\cite{bru16,lud13} that for the case of time-independent coupling, such a set of definitions may be obtained by defining a system energy that consists of the expectation value of the dot Hamiltonian with half the coupling Hamiltonian added,
\begin{align}
\label{systemenergy}
E(t)=\left\langle H_\text{D}(t)+\frac{1}{2}H_\text{T}(t)\right\rangle.
\end{align}
This kind of splitting is indicative of the fact that in the presence of strong coupling, one cannot simply identify the dot with the ``system'', and the lead with the ``bath'', in the thermodynamic sense of these terms. Moreover, energy added to the total ensemble by the coupling drive must be distributed between system and bath. In the following, we show that this choice of system energy remains valid in the case of driven coupling.

The expectation values in Eq.~\eqref{systemenergy} are readily expressed in terms of Green's functions, leading to the exact result
\begin{align}
\label{systemenergy_exact}
E(t)&=\int\frac{\text{d}\omega}{2\pi}\omega f(\omega) A(t,\omega)\notag\\
&-2\int\frac{\text{d}\omega}{2\pi}f(\omega) \Im \left[\partial_t V(t,\omega)V^*(t,\omega)\right],
\end{align}
with \(V(t,\omega)\) and \(A(t,\omega)\) as in Eqs.~\eqref{vdef} and \eqref{adef}. We define the rate of change in work performed on the system as split into three parts,
\begin{align}
\label{work}
\dot{W}(t)\equiv\dot{W}_\text{SB}(t)+\dot{W}_\text{C}(t)+\dot{W}_{\text{B}}(t),
\end{align}
where \(\dot{W}_\text{SB}(t)=\left\langle\partial_t H(t)\right\rangle\) and \(\dot{W}_\text{C}(t)=\mu\partial_t N(t)\) denote the power supplied by the drive to system and bath, and the chemical work rate associated with particle flow into the dot, respectively. Below, we will always use the dot symbol to denote a rate, as opposed to \(\partial_t\) which stands for a time derivative.
\(\dot{W}_\text{SB}(t)\) is found to be
\begin{align}
\label{work_SB}
\dot{W}_\text{SB}(t)&=\partial_t\epsilon(t)N(t)+\langle\partial_t H_\text{T}(t)\rangle\notag\\
&=\partial_t\epsilon(t)\int\frac{\text{d}\omega}{2\pi}f(\omega)A(t,\omega)\notag\\
&+\frac{2\partial_t\Gamma(t)}{\sqrt{\Gamma(t)}}\int\frac{\text{d}\omega}{2\pi}f(\omega)\Im{V(t,\omega)},
\end{align}
whereas the chemical chemical work rate equals
\begin{align}
\dot{W}_\text{C}(t)&=4\mu\sqrt{\Gamma(t)}\int\frac{\text{d}\omega}{2\pi}f(\omega)\Re{V(t,\omega)}\notag\\
&-2\mu\Gamma(t)\int\frac{\text{d}\omega}{2\pi}f(\omega)A(t,\omega).
\end{align}
The third term in Eq.~\eqref{work} is a work done by the system-bath coupling to change the particle numbers in the bath. It is reminiscent of the work that the coupling needs to do to create a volume in the bath recently identified in Ref.~[\onlinecite{jar17}], but in the grand canonical ensemble. In order to obtain the rate of work performed on the system only, this contribution therefore needs to be subtracted. It is given by
\begin{align}
\label{wb}
\dot{W}_\text{B}(t) \equiv \frac{1}{\pi}\partial_t\Gamma(t)= 2\partial_t \Gamma(t) \partial_\mu N^\text{eq}_{\text{B}}/\rho_0,
\end{align}
where $\rho_0$ is the lead density of states as in Sec.~\ref{sec_model} and \(N^\text{eq}_{\text{B}}=\rho_0 \int\frac{\text{d}\omega}{2\pi}f(\omega)\) denotes the (infinite) equilibrium particle number in the lead. 
Its change with respect to the chemical potential, \(\partial_\mu N^\text{eq}_{\text{B}}\), is finite and can be seen as the grand canonical analogue of a compressibility.  
One sees therefore that \(\dot{W}_\text{B}\) arises from changes in the tunneling linewidth \(\Gamma(t)\) which modify the level repulsion among the levels in the lead, in turn causing a change in the lead particle number. 

Among the results of Sec.~\ref{sec_secondlaw}, we will find that this definition of \(\dot{W}_\text{B}\) is compatible with the second law of thermodynamics. Lastly, we define the heat current flowing into the system as
\begin{align}
\label{heat}
\dot{Q}(t)=-\partial_t\left\langle H_\text{B}+\frac{1}{2}H_\text{T}(t)\right\rangle-\dot{W}_\text{C}(t)-\dot{W}_\text{B}(t).
\end{align}
Mirroring Eq.~\eqref{systemenergy}, this definition associates half of the coupling energy with the bath, and explicitly features the reversed work flows due to particle transfer and work performed on the bath.

If we sum up the definitions from Eqs.~\eqref{systemenergy}, \eqref{work}, and \eqref{heat}, we obtain the energy balance
\begin{align}
\label{firstlaw}
\partial_t E=\dot{Q}+\dot{W},
\end{align}
which makes manifest the first law of thermodynamics in our system, with \(E\) taking on the role of the internal energy.

\section{Link to equilibrium}
\label{sec_equilibrium}
In this section we start by documenting an alternative approach to the adiabatic limit of the model. Then we take the limit of infinitely slow drive, \(\partial_t\epsilon\to0\) and \(\partial_t\Gamma\to0\) of the definitions made in Sec.~\ref{sec_firstlaw} and compare the two sets of findings, thus ensuring that our quantities reduce to the correct adiabatic limit.

The starting point for this equilibrium discussion is a grand canonical ensemble of the ``super-system'', which comprises both dot and lead. The latter are assumed to be weakly coupled to a ``super-bath'' characterized by an inverse temperature \(\beta\) and chemical potential \(\mu\). This setup has been employed for a classical model,\cite{str17} as well as for the resonant level model with a single drive parameter.\cite{bru16}

In the absence of drive, and assuming that the super-system is coupled to the super-bath by energy and particle exchange, we can obtain the weak coupling thermodynamics of the super-system from the equilibrium grand canonical potential
\begin{align}
\label{grandcan}
\Omega^\text{eq}&\equiv-\frac{1}{\beta}\log{\tr{\text{e}^{-\beta(H-\mu\mathcal{N})}}}\notag\\
&=-\frac{1}{\beta}\int\frac{d\omega}{2\pi}\rho(\omega)\log{\left(1+\text{e}^{-\beta(\omega-\mu)}\right)}
\end{align}
where $\mathcal{N}$ is the particle number operator of the super-system. Here, \(\rho(\omega)\) denotes the stationary density of states of the super-system. It is defined as the sum of dot and lead contributions, which in terms of Green's functions is given by
\begin{align}
\label{adiabaticdos}
\rho(\omega)=-2\Im{D}^\text{R}(\omega)-2\sum_{k}\Im{G}_{k}^\text{R}(\omega).
\end{align}
Following Ref.~[\onlinecite{bru16}], we note that the sum over the exact lead Green's functions obeys the Dyson equation
\begin{align}
\sum_{k}G_{k}^\text{R}(\omega)=\sum_{k}G_{0,k}^{\text{R}}(\omega)+|\gamma|^2D^\text{R}(\omega)\sum_{k}\left[G_{0,k}^{\text{R}}(\omega)\right]^2.
\end{align}
Writing the unperturbed lead Green's function in the frequency domain, \(G_{0,k}^{\text{R}}(\omega)=(\omega-\epsilon_k+\ci 0^+)^{-1}\), we see that the correction becomes
\begin{align}
&|\gamma|^2 D^\text{R}(\omega) \sum_{k}\left[G_{0,k}^{\text{R}}(\omega)\right]^2\notag\\
&=-D^\text{R}(\omega) \partial_\omega\left[|\gamma|^2\sum_{k}G_{0,k}^{\text{R}}(\omega)\right]
=-D^\text{R}(\omega) \partial_\omega \Sigma^R(\omega)
\end{align}
According to Eq.~(\ref{sigmaret}), this correction vanishes in the wide-band limit. Therefore the lead component of \(\rho(\omega)\) is not renormalized by the coupling, facilitating its interpretation as a ``pure bath'' term which does not contribute to the system dynamics and can be dropped from Eq.~\eqref{grandcan}. We thus define the system grand canonical potential of mean force,
\begin{align}
\label{systemgrandcan}
\Omega^\text{eq}_\text{S}&\equiv-\frac{1}{\beta}\int\frac{d\omega}{2\pi}A_0(\omega)\log{\left(1+\text{e}^{-\beta(\omega-\mu)}\right)},
\end{align}
where \(A_0(\omega)\) is the stationary spectral function defined in Eq.~\eqref{A0def}.
Changes in \(\Omega^\text{eq}_\text{S}\) resulting from modification of the system parameters are equal to the corresponding changes in \(\Omega^\text{eq}\). With this choice of \(\Omega^\text{eq}_\text{S}\), we can use equilibrium thermodynamics to obtain expressions for the equilibrium values of particle number, system entropy and energy,
\begin{align}
N^\text{eq}&=-\partial_\mu\Omega^\text{eq}_\text{S}=\int\frac{\text{d}\omega}{2\pi}A_0(\omega)f(\omega)\label{eqdef_N}\\
S^\text{eq}&=-\partial_T\Omega^\text{eq}_\text{S}=\int\frac{\text{d}\omega}{2\pi}A_0(\omega)\sigma_f(\omega)\label{eqdef_S}\\
E^\text{eq}&=\Omega^\text{eq}_\text{S}+\mu N^\text{eq}+\frac{1}{\beta}S^\text{eq}=\int\frac{\text{d}\omega}{2\pi}A_0(\omega)\omega f(\omega)\label{eqdef_E}.
\end{align}
where in Eq.~\eqref{eqdef_S} we defined the frequency-resolved entropy factor
\begin{align}
\sigma_f(\omega)=-f(\omega)\log{f(\omega)}-[1-f(\omega)]\log{[1-f(\omega)]}.
\end{align}
These quantities can be related to the adiabatic limit of the definitions in Sec.~\ref{sec_firstlaw} by introducing a parametric time dependence in \(A_0\) via \(\epsilon(t)\) and \(\Gamma(t)\). The expressions obtained by substituting this time-dependent Lorentzian \(A_0(t,\omega)\) into Eqs.~\eqref{eqdef_N}, \eqref{eqdef_S}, and \eqref{eqdef_E} can then be used to calculate the adiabatic particle and energy currents,
\begin{align}
\partial_t N^\text{eq}(t)&=\int\frac{\text{d}\omega}{2\pi}f\left(\partial_\Gamma A_0\partial_t\Gamma+\partial_\epsilon A_0\partial_t\epsilon\right)=[\partial_t N]^{(1)}\notag\\
\partial_t E^\text{eq}(t)&=\int\frac{\text{d}\omega}{2\pi}\omega f\left(\partial_\Gamma A_0 \partial_t \Gamma+\partial_\epsilon A_0 \partial_t \epsilon\right)=[\partial_t E]^{(1)},
\end{align}
where the right hand sides refer to the adiabatic expansions of the time derivatives of Eqs.~\eqref{number} and \eqref{systemenergy_exact}, respectively, which are detailed in App.~\ref{app_expansion}. Similarly, we can interpret the time derivative of the grand canonical potential as the rate of mechanical work performed on the system,
\begin{align}
\label{eqwork}
\partial_t \Omega^\text{eq}_\text{S}(t)&=-\frac{1}{\beta}\int\frac{\text{d}\omega}{2\pi}\left(\partial_\Gamma A_0 \partial_t \Gamma+\partial_\epsilon A_0 \partial_t \epsilon\right)\notag\\
&\qquad\qquad\times\left[\sigma_f(\omega)-\beta(\omega-\mu)f(\omega)\right]\notag\\
&=\dot{W}^{(1)}_\text{SB}+\dot{W}_\text{B},
\end{align}
as can be gleaned from Eq.~\eqref{heatexp1}. Hence we conclude that the adiabatic limit of the system quantities defined in Sec.~\ref{sec_firstlaw} matches the result of adiabatic weak coupling thermodynamics as encoded in the grand canonical potential from Eq.~\eqref{systemgrandcan}.

\section{Second law}
\label{sec_secondlaw}
Having obtained a definition for the heat flowing into the system, we now address the question of how to define the system entropy. To this end, we generalize the adiabatic expression given in Eq.~\eqref{eqdef_S} and show the compatibility of this choice with the previous definitions by exhibiting the second law.

In analogy to the non-adiabatic result for the particle number on the dot, Eq.~\eqref{number}, we define the system entropy beyond the adiabatic limit by replacing the Lorentzian spectral function \(A_0\) in Eq.~\eqref{eqdef_S} with the function \(A\) as in Sec.~\ref{sec_model},
\begin{align}
\label{entropy}
S(t)\equiv\int\frac{\text{d}\omega}{2\pi}A(t,\omega)\sigma_f(\omega).
\end{align}
Next, we show the second law in the sense that the entropy production rate is non-negative up to second order in the quasi-adiabatic expansion:
\begin{align}\label{secondlaw}\partial_t S-\beta\dot{Q}\geq0,\end{align}
with equality up to first order. Details of the expansion can be found in App.~\ref{app_expansion}.

Starting with the first order, we note that on the one hand, the derivative of Eq.~\eqref{entropy} is approximated by
\begin{align}
\label{entropychange}
\left[\partial_t S\right]^{(1)}=\int\frac{\text{d}\omega}{2\pi}\sigma_f(\omega)\left(\partial_\Gamma A_0\partial_t\Gamma+ \partial_\epsilon A_0\partial_t\epsilon\right),
\end{align}
whereas on the other hand, we obtain the corresponding terms for the heat flow from Eq.~\eqref{heatexp1},
\begin{align}
\dot{Q}^{(1)}&=\frac{1}{\beta}\int\frac{\text{d}\omega}{2\pi}\sigma_f\left(\partial_\Gamma A_0\partial_t\Gamma+ \partial_\epsilon A_0\partial_t\epsilon\right)\notag\\
&-\frac{1}{\beta}\int\frac{\text{d}\omega}{2\pi}\log{\left(1+\text{e}^{-\beta\omega}\right)}\partial_\omega\left(-\frac{\omega-\epsilon}{\Gamma}A_0\partial_t\Gamma-A_0\partial_t\epsilon\right)\notag\\
&-\int\frac{\text{d}\omega}{2\pi}fA_0\left(\frac{\omega-\epsilon}{\Gamma}\partial_t\Gamma+\partial_t\epsilon\right)-\dot{W}_\text{B}.
\end{align}
An integration by parts in the second integral yields a term that cancels the third integral, as well as a boundary contribution equal to \(\partial_t\Gamma/\pi\), thus canceling \(-\dot{W}_\text{B}\). The first integral coincides with \(\left[\partial_t S\right]^{(1)}\), which implies that to first order in adiabatic expansion, the change in system entropy is entirely due to heat flow,
\begin{align}
\left[\partial_t S\right]^{(1)}=\beta\dot{Q}^{(1)}.
\end{align}
Moreover, the quasi-adiabatic expansion shows that our definitions give rise to exact differentials for the system energy as well as reversible work and heat flows to first order in time derivatives,
\begin{align}
\partial_\Gamma\partial_\epsilon\dot{Q}^{(1)}&=\partial_\epsilon\partial_\Gamma\dot{Q}^{(1)}, \notag \\ \partial_\Gamma\partial_\epsilon\dot{W}^{(1)}&=\partial_\epsilon\partial_\Gamma\dot{W}^{(1)}.
\end{align}
This should be contrasted with the observation made in Ref.~[\onlinecite{gal15}], that a splitting of the coupling energy as in Eq.~\eqref{systemenergy} makes a certain class of heat definitions problematic in this regard as soon as driven coupling is considered. As our choice of \(\dot{Q}\) from Eq.~\eqref{heat} differs from the class of heat definitions considered in Ref.~[\onlinecite{gal15}], this problem does not arise here.

Moving to the second order in time derivatives, we note that the relevant part of \(\dot{Q}^{(2)}\) are given by the last two integrals in Eq.~\eqref{heatexp},
\begin{align}
\label{Q2}
\delta\dot{Q}^{(2)}&\equiv\dot{Q}^{(2)}-\dot{Q}^{(1)}\notag\\
&=\frac{\Gamma^2}{2}\int\frac{\text{d}\omega}{2\pi}\partial_\omega f A_0^2\left(\partial_t\frac{\omega-\epsilon}{\Gamma}\right)^2\notag\\
&+\int\frac{\text{d}\omega}{2\pi}(\omega-\mu)\partial_\omega f\partial_t\left(\frac{A_0^2\Gamma}{2}\partial_t\frac{\omega-\epsilon}{\Gamma}\right).
\end{align}
On the other hand, the expansion of \(\partial_t S\) is immediate from Eq.~\eqref{Aexp},
\begin{align}
\delta\dot{S}^{(2)}&\equiv\left[\partial_t S\right]^{(2)}-\left[\partial_t S\right]^{(1)}\notag\\
&=\int\frac{\text{d}\omega}{2\pi}\partial_\omega\sigma_f\partial_t\left(\frac{A_0^2\Gamma}{2}\partial_t\frac{\omega-\epsilon}{\Gamma}\right).
\end{align}
Using \(\partial_\omega\sigma_f(\omega)=\beta(\omega-\mu)\partial_\omega f\), we see that this matches the second term in Eq.~\eqref{Q2}, and therefore we obtain
\begin{align}
\label{secondlaw2}
\delta\dot{S}^{(2)}-\beta\delta\dot{Q}^{(2)}=-\frac{\Gamma^2}{2}\int\frac{\text{d}\omega}{2\pi}(\partial_\omega f) A_0^2\left(\partial_t\frac{\omega-\epsilon}{\Gamma}\right)^2\geq0,
\end{align}
which proves the second law of thermodynamics~\eqref{secondlaw} to second order. We remark that the integral occurring in Eq.~\eqref{secondlaw2} equals the negative of the second-order term in the work performed on the super-system, Eq.~\eqref{superworkexp}. Hence the excess entropy production beyond the adiabatic limit can be interpreted as a consequence of mechanical friction causing heat to leave the system.

\section{Comparison with exact numerical results}
\label{sec_numerics}
In this section, we compare the analytical results which were derived up to the second order in the drive speed with exact numerical results. For this purpose, we study a protocol where both dot and coupling drives are cosine-shaped,
\begin{align}
\epsilon(t)&=\epsilon_0+\Delta_\epsilon\cos{\omega_\epsilon t}\notag\\
\Gamma(t)&=\Gamma_0+\Delta_\Gamma\cos{\omega_\Gamma t},
\end{align}
and we set \(\mu=0\). By tuning the parameters, this protocol can be made to include the regimes of strong dot-lead coupling and non-adiabatic drive.

\begin{figure}[!t]
     \includegraphics[width=0.99\columnwidth]{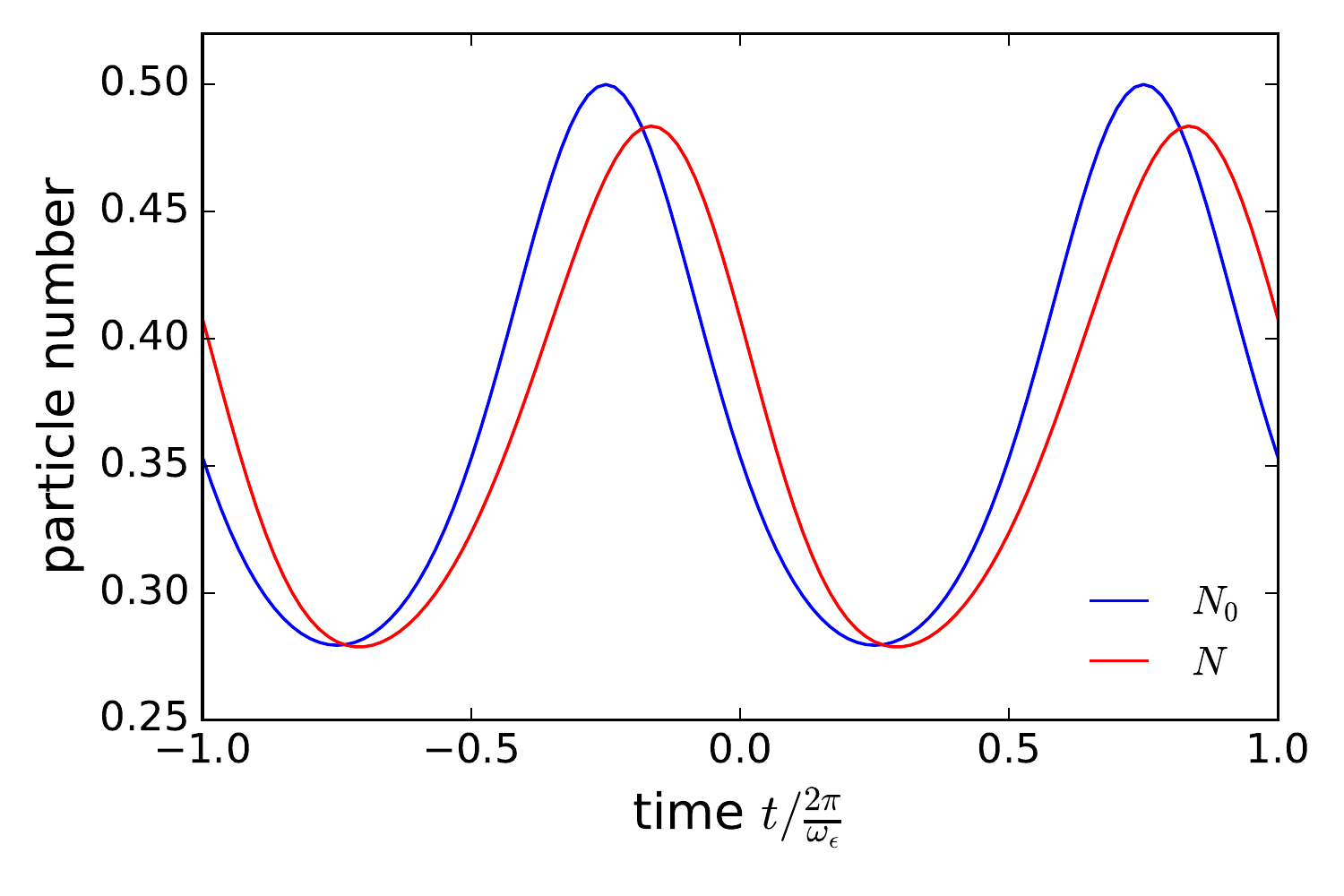}
 \caption{Dot particle number as a function of time, for \(\epsilon_0=0.5\), \(\Delta_\epsilon=0.5\), \(\omega_\epsilon=0.5\), \(\Gamma_0=1\), \(\Delta_\Gamma=0.2\), \(\omega_\Gamma=0.5\). \emph{Blue}: adiabatic \(N_0(t)\) obtained using the Lorentzian spectral density \(A_0(t,\omega)\). \emph{Red}: Exact \(N(t)\) from Eq.~\eqref{number}. }
\label{fig_number}
\end{figure}

The dot particle number \(N(t)\) as calculated from Eq.~\eqref{number} is displayed in Fig.~\ref{fig_number}, and contrasted with the adiabatic result for \(N(t)\) that is obtained by using the Lorentzian spectral function \(A_0(t,\omega)\), with time-dependence parameters $\epsilon(t)$ and $\Gamma(t)$. We observe that non-adiabaticity causes the exact result to lag behind the adiabatic one, in line with the retarded character of the time integrals in the definition of \(A(t,\omega)\).

\begin{figure}[!t]
     \includegraphics[width=0.99\columnwidth]{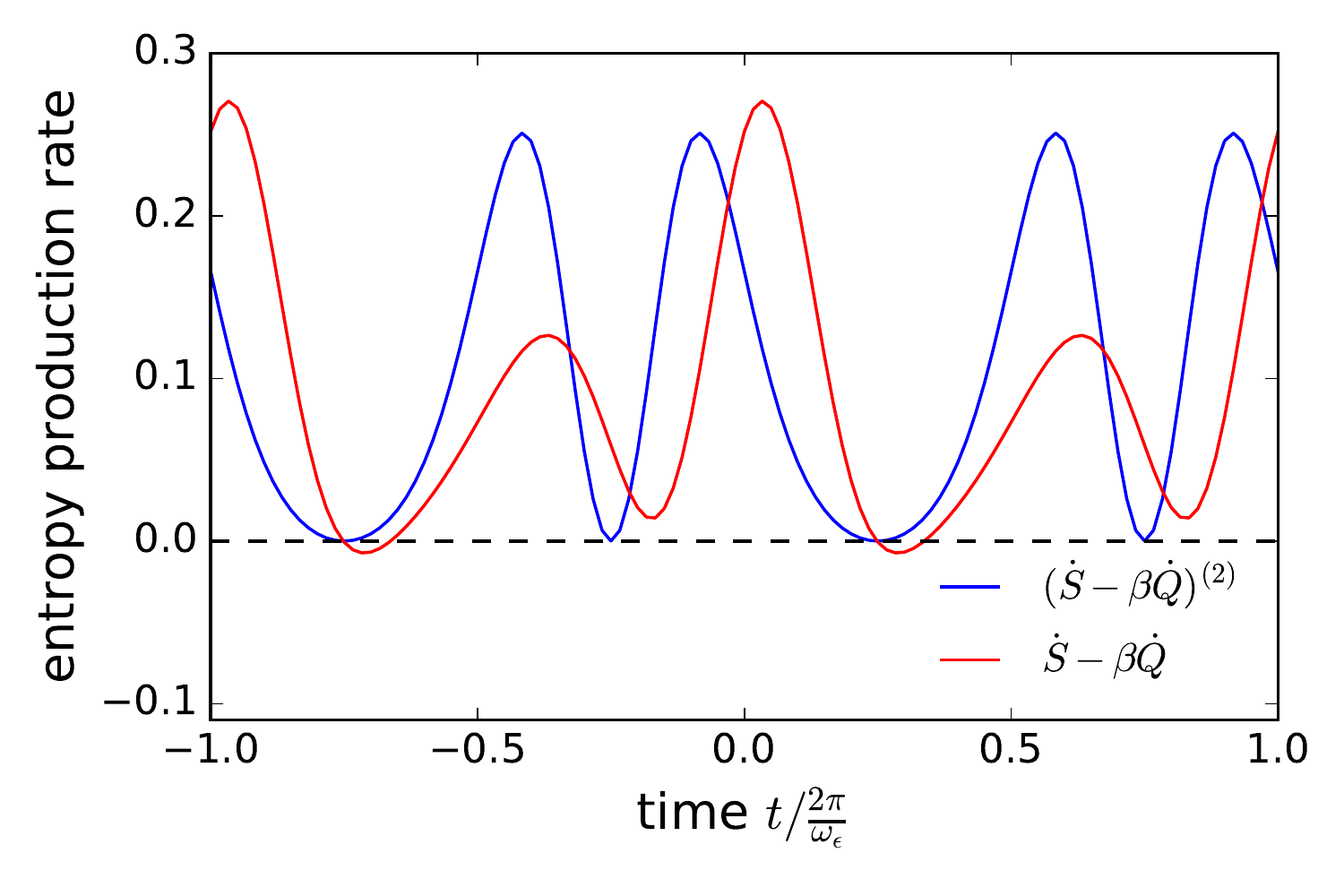}
 \caption{Difference of entropy production rate \(\partial_t S(t)\) and inverse temperature multiplied by heat, for \(\epsilon_0=0.5\), \(\Delta_\epsilon=0.5\), \(\omega_\epsilon=0.5\), \(\Gamma_0=1\), \(\Delta_\Gamma=0.2\), \(\omega_\Gamma=0.5\). \emph{Blue}: Second-order quasi-adiabatic entropy production rate \((\partial_t S-\beta\dot{Q})^{(2)}\). This rate is positive for all times, in accordance with the second law as in Eq.~\eqref{secondlaw2}. \emph{Red}: \(\partial_t S-\beta\dot{Q}\) as calculated from Eqs.~\eqref{heat} and \eqref{entropy}. This rate can become negative beyond the adiabatic limit, reflecting the influence of higher orders in the quasi-adiabatic expansion. Moreover, the non-Markovianity of the system itself may lead to negative transients. }
\label{fig_entropy}
\end{figure}

Moreover, in Fig.~\ref{fig_entropy} we compare the exact entropy production \(\partial_t S-\beta\dot{Q} \), calculated numerically based on Eqs.~(\ref{entropy}) and (\ref{heat}) with the corresponding result (\ref{secondlaw2}) in the quasi-adiabatic limit. While the exact result indeed converges to the quasi-adiabatic case for slow driving, significant deviations from the quasi-adiabatic result occur already for parameters where $N(t)$ is still very close to the adiabatic result. In particular, whereas the time integral of the entropy production rate over a drive cycle is positive, the rate itself features negative transients, which are a sign of the non-Markovianity inherent in our model.\cite{breuer_petruccione_book}

\section{Conclusions}
\label{sec_discussion}
We have presented an analytical solution of the resonant level model in the wide-band limit, in the presence of both time-dependent dot energy and tunnel coupling. We defined thermodynamic quantities, which we calculated using this solution, and found them to be in accordance with the first law of thermodynamics. We found that the adiabatic limit of our definitions matches the results known from stationary thermodynamics. Finally, a quasi-adiabatic expansion allowed us to verify the second law of thermodynamics to second order in time derivatives of the drive protocol.

It is worthwhile to compare the definitions made here with other recent research on the subject. The choice of a system energy that includes half of the coupling contribution has previously been studied in the case of constant tunnel coupling.\cite{lud13,bru16,och16} Our work generalizes these results to driven coupling, whereby we find that the terms \(\pm\dot{W}_\text{B}\) need to be added to the definitions \eqref{work} and \eqref{heat} of the rate of work performed by the system and the heat current flowing through it, respectively. Specifically, the quasi-adiabatic expansion of our exact results matches the findings of Ref.~[\onlinecite{bru16}] if we take the limit of time-independent $\Gamma$. Our definitions also give rise to a state function for the reversible heat, which resolves the issue pointed out in Ref.~[\onlinecite{gal15}] for time-dependent \(\Gamma\).

Compared to the case of constant tunneling, time-dependent tunnel amplitudes also give rise to a nontrivial gauge invariance. A time-dependent dot level energy $\epsilon(t)$ can easily be mapped onto a time-dependent tunnel amplitude, $\gamma(t) \propto \gamma_0(t) \exp[-i\int^t \text{d}s \epsilon(s)]$. Our results all have this gauge invariance. In contrast, several works have suggested that system quantities can be defined by isolating $\epsilon(t)$ dependent quantities in the ``super-system'' observables.\cite{bru16,och16} However, this procedure is not gauge-invariant and thus cannot be used in the case of time-dependent tunnel amplitudes.

Several challenges remain to be overcome on the way to a full understanding of non-equilibrium quantum thermodynamics in the resonant level model: Beyond the wide-band approximation, the precise correspondence between the Green's function approach and the results obtained for the grand canonical ensemble in the adiabatic limit as in Sec.~\ref{sec_equilibrium} is still unclear. Similar problems arise if one considers higher moments of Hamiltonians instead of expectation values only.\cite{och16} Finally, it is an appealing prospect to find a version of the second law that holds for all orders in drive speed as opposed to just second order.

\begin{acknowledgments}
TLS and PH acknowledge support by the National Research Fund, Luxembourg under grant ATTRACT 7556175. ME acknowledges support by the European Research Council project NanoThermo (ERC-2015-CoG Agreement No. 681456).
\end{acknowledgments}

\appendix
\section{Adiabatic limit}
\label{app_limit}
In the following, we derive approximations for the quantities defined in Sec.~\ref{sec_firstlaw} for the case of slow driving. First, we establish the adiabatic limit of infinitely slow drive. Then, we move to the quasi-adiabatic case by expanding the exact expressions in terms of time derivatives of the drive protocol \([\epsilon(t),\Gamma(t)]\).

The function
\begin{align}
V(t,\omega)=\int_{-\infty}^t \text{d}s \sqrt{\Gamma(s)} \text{e}^{\int_s^t \text{d}y \left[\ci \omega-\ci \epsilon(y)-\Gamma(y)\right]}
\end{align}
from Eq.~\eqref{vdef} is the central subject of the calculations in this section. Its static limit is obtained by assuming constant \(\epsilon\equiv\epsilon_0\) and \(\Gamma\equiv\Gamma_0\),
\begin{align}
V^{(0)}(\omega)=\frac{\sqrt{\Gamma_0}}{\ci (\epsilon_0-\omega)+\Gamma_0}.
\end{align}
Using this to calculate \(A^{(0)}=2|V^{(0)}(\omega)|^2\), we obtain
\begin{align}
A^{(0)}(\omega)=\frac{2\Gamma_0}{(\omega - \epsilon_0)^2 + \Gamma_0^2},
\end{align}
which coincides with the spectral function \(A_0(\omega)\). From Eq.~\eqref{number}, we immediately obtain the particle number,
\begin{align}
N^{(0)}=\int\frac{\text{d}\omega}{2\pi}A_0(\omega)f(\omega)=\int\frac{\text{d}\omega}{2\pi}\frac{2\Gamma_0}{(\omega - \epsilon_0)^2 + \Gamma_0^2}f(\omega).
\end{align}
Similarly, we obtain for the system energy,
\begin{align}
E^{(0)}&=\int\frac{\text{d}\omega}{2\pi}\omega f(\omega) A_0(\omega),
\end{align}
since the second term in Eq.~\eqref{systemenergy_exact} is approximated by zero.

\section{Quasi-adiabatic expansion}
\label{app_expansion}
In this section, we move beyond the adiabatic limit in approximating the particle number, system energy, as well as heat and work rates. To this end we reinstate the time dependence of \(\Gamma\) and \(\epsilon\) in \(V(t,\omega)\) which occurs in all the quantities considered here. We then expand both drives up to second order in time derivatives resulting in the expansion
\begin{widetext}
\begin{align}
\label{vexp}
V^{(2)}(t,\omega)&=\frac{\sqrt{\Gamma}}{\ci  (\epsilon - \omega) + \Gamma}
-
    \frac{\partial_t\Gamma}{2 \sqrt{\Gamma}} \frac{1}{[\ci  (\epsilon - \omega) + \Gamma]^2}
+
     \left[ \frac{\partial_t^2\Gamma}{4 \sqrt{\Gamma}} - \frac{(\partial_t\Gamma)^2}{8 \sqrt{\Gamma}^3} + \frac{\sqrt{\Gamma}}{2} (\ci  \partial_t\epsilon + \partial_t\Gamma)  \right]  \frac{2}{[\ci  (\epsilon - \omega) + \Gamma]^3} \notag \\
&+
    \left[ - \frac{\partial_t\Gamma (\ci \partial_t\epsilon + \partial_t\Gamma)}{4 \sqrt{\Gamma}} - \frac{1}{6} \sqrt{\Gamma} (\ci  \partial_t^2\epsilon + \partial_t^2\Gamma)\right]\frac{6}{[\ci  (\epsilon - \omega) + \Gamma]^4}
+
    \frac{3 \sqrt{\Gamma} (\ci  \partial_t\epsilon + \partial_t\Gamma)^2}{[\ci  (\epsilon - \omega) + \Gamma]^5},
\end{align}
\end{widetext}
where all drives are evaluated at time \(t\). By substituting this expression, we readily obtain second-order results for the currents \(\partial_t N\), \(\dot{W}\), \(\partial_t E\), and \(\dot{Q}\). Since the second-order contributions to these quantities go beyond the adiabatic results in the sense of Sec.~\ref{sec_equilibrium}, we refer to them as quasi-adiabatic expansion.

The particle current is given by the time derivative of Eq.~\eqref{number}.
\begin{align}
\partial_t N(t)&=\int\frac{\text{d}\omega}{2\pi}f(\omega)\partial_t A(t,\omega),
\end{align}
where by substituting Eq.~\eqref{vexp}, we find the second-order expansion of \(\partial_t A(t,\omega)=2\partial_t|V(t,\omega)|^2\) to be given by
\begin{align}
\label{Aexp}
\left[\partial_t A(t,\omega)\right]^{(2)}&=\left(\partial_\Gamma A_0\partial_t\Gamma+\partial_\epsilon A_0\partial_t\epsilon\right)\notag\\
&-\partial_t\partial_\omega\left(\frac{A_0^2\Gamma}{2}\partial_t\frac{\omega-\epsilon}{\Gamma}\right),
\end{align}
where we suppress the arguments \(t\) and \(\omega\) from here onward. Therefore, the second-order quasi-adiabatic expansion of the particle current reads
\begin{align}
\label{numberexp}
\left[\partial_t N(t)\right]^{(2)}&=\int\frac{\text{d}\omega}{2\pi}f\left(\partial_\Gamma A_0\partial_t\Gamma+\partial_\epsilon A_0\partial_t\epsilon\right)\notag\\
&+\int\frac{\text{d}\omega}{2\pi}\partial_\omega f\partial_t\left(\frac{A_0^2\Gamma}{2}\partial_t\frac{\omega-\epsilon}{\Gamma}\right).
\end{align}
Analogously, by starting from Eq.~\eqref{systemenergy_exact}, we find the second-order expression for the system energy current,
\begin{align}
\label{systemenergyexp}
\left[\partial_t E_\text{S}\right]^{(2)}
&=\int\frac{\text{d}\omega}{2\pi}\omega f\left(\partial_\Gamma A_0 \partial_t \Gamma+\partial_\epsilon A_0 \partial_t \epsilon\right)\notag\\
&+\int\frac{\text{d}\omega}{2\pi}\omega\partial_\omega f\partial_t\left(\frac{A_0^2\Gamma}{2}\partial_t\frac{\omega-\epsilon}{\Gamma}\right).
\end{align}
The work flow into the system consists of three distinct contributions, \(\dot{W}=\dot{W}_\text{SB}+\dot{W}_\text{C}+\dot{W}_{\text{B}}\), the first two of which require expansion: The power applied to the super-system is approximated by expanding Eq.~\eqref{work_SB},
\begin{align}
\label{superworkexp}
\dot{W}_\text{SB}^{(2)}&=\int\frac{\text{d}\omega}{2\pi} fA_0\left(\partial_t\epsilon+\frac{\omega-\epsilon}{\Gamma}\partial_t\Gamma\right)\notag\\
&-\frac{\Gamma^2}{2}\int\frac{\text{d}\omega}{2\pi}\partial_\omega fA_0^2\left(\partial_t\frac{\omega-\epsilon}{\Gamma}\right)^2,
\end{align}
and the chemical work flow is given by \(\dot{W}_\text{C}^{(2)}=\mu[\partial_t N]^{(2)}\), which is immediate from Eq.~\eqref{numberexp}. Finally, we consider the heat current, \(\dot{Q}=-\partial_t\left\langle H_\text{B}+H_\text{T}/2\right\rangle-\dot{W}_\text{C}-\dot{W}_\text{B}\). The last term is the reverse of the first-order expression quantifying mechanical work performed on the bath, whereas the first two terms can be expressed as as
\begin{align}
&-\partial_t\left\langle H_\text{B}+H_\text{T}/2\right\rangle-\mu\partial_t N=-\dot{W}_\text{SB}+\partial_t E-\mu\partial_t N,
\end{align}
which leads to the approximation
\begin{align}
\label{heatexp}
\dot{Q}^{(2)}&=-\dot{W}_\text{SB}^{(2)}+\left(\partial_t E\right)^{(2)}-\mu\partial_t N^{(2)}-\dot{W}_\text{B}\notag\\
&=-\int\frac{\text{d}\omega}{2\pi}fA_0\left(\partial_t\epsilon+\frac{\omega-\epsilon}{\Gamma}\partial_t \Gamma\right)\notag\\
&+\int\frac{\text{d}\omega}{2\pi}(\omega-\mu)f\left(\partial_\Gamma A_0\partial_t\Gamma+ \partial_\epsilon A_0\partial_t\epsilon\right)\notag\\
&+\frac{\Gamma^2}{2}\int\frac{\text{d}\omega}{2\pi}\partial_\omega f A_0^2\left(\partial_t\frac{\omega-\epsilon}{\Gamma}\right)^2\notag\\
&+\int\frac{\text{d}\omega}{2\pi}(\omega-\mu)\partial_\omega f\partial_t\left(\frac{A_0^2\Gamma}{2}\partial_t\frac{\omega-\epsilon}{\Gamma}\right)-\dot{W}_\text{B}.
\end{align}

Making use of the relations \(\beta(\omega-\mu) f(\omega)=\sigma_f(\omega)-\log{\left(1+\text{e}^{-\beta\omega}\right)}\) and \(\partial_\Gamma A_0(\omega)=-\partial_\omega[(\omega-\epsilon)A_0/\Gamma]\), we can rewrite the first-order terms as
\begin{align}
\label{heatexp1}
\dot{Q}^{(1)}&=\frac{1}{\beta}\int\frac{\text{d}\omega}{2\pi}\sigma_f\left(\partial_\Gamma A_0\partial_t\Gamma+ \partial_\epsilon A_0\partial_t\epsilon\right)\notag\\
&-\frac{1}{\beta}\int\frac{\text{d}\omega}{2\pi}\log{\left(1+\text{e}^{-\beta\omega}\right)}\partial_\omega\left(-\frac{\omega-\epsilon}{\Gamma}A_0\partial_t\Gamma-A_0\partial_t\epsilon\right)\notag\\
&-\int\frac{\text{d}\omega}{2\pi}fA_0\left(\frac{\omega-\epsilon}{\Gamma}\partial_t\Gamma+\partial_t\epsilon\right)-\dot{W}_\text{B},
\end{align}
where after integrating by parts, the second line cancels the third.
Furthermore, by comparing Eq.~\eqref{heatexp1} to the time derivative of the equilibrium grand canonical potential of Eq.~\eqref{systemgrandcan} and the work rate in Eq.~\eqref{superworkexp} we obtain the relation \eqref{eqwork}.

\bibliography{thermo}

\end{document}